\documentclass[aps,prb,amsmath,reprint,superscriptaddress]{revtex4-1}
\usepackage{amsmath,txfonts,bm,dcolumn,graphicx,physics}
\begin{document}

\title{On-the-fly CASPT2 Surface Hopping Dynamics} 
\author{Jae Woo Park}
\email{jwpk1201@northwestern.edu}
\author{Toru Shiozaki}
\affiliation{Department of Chemistry, Northwestern University, 2145 Sheridan Rd., Evanston, IL 60208, USA.}
\date{\today}

\begin{abstract}
We report the development of programs for on-the-fly surface hopping dynamics simulations in the gas and condensed phases on the potential energy surfaces computed by 
multistate multireference perturbation theory (XMS-CASPT2) with full internal contraction.
On-the-fly nonadiabatic dynamics simulations are made possible by improving the algorithm for XMS-CASPT2 nuclear energy gradient and derivative coupling evaluation.
The program is interfaced to a surface hopping dynamics program, {\sc Newton-X}, and a classical molecular dynamics package, {\sc tinker}, to realize such simulations.
On-the-fly XMS-CASPT2 surface-hopping dynamics simulations of 9H-adenine and an anionic GFP model chromophore (\textit{para}-hydroxybenzilideneimidazolin-5-one) in water
are presented to demonstrate the applicability of our program to sizable systems. 
Our program is implemented in the {\sc bagel} package, which is publicly available under the GNU General Public License.
\end{abstract}

\maketitle

\section{Introduction}

Photochemical reactions involving multiple electronic states in condensed phases are often studied by means of first-principles nonadiabatic dynamics simulations\cite{Tully1990JCP,HammesSchiffer1994JCP,Levine2007ARPC,Barbatti2011WIREs,Gonzalez2012CPC,Wang2016JPCL,Subotnik2016ARPC} combined with molecular mechanics
for the environment (QM/MM).
In such simulations, especially when nonradiative decays to the ground state is important, the complete active space self-consistent field (CASSCF) method is often used
to calculate the potential energy surfaces of the ground and excited states of molecules.
One may also use semiempirical methods,
such as floating occupation molecular orbital-configuration interaction (FOMO-CI)\cite{Toniolo2003JPCA,Toniolo2004Faraday} and orthogonalization-corrected
methods combined with multireference configuration interaction (OM\textit{x}/MRCI).\cite{Koslowski2003JCC,Fabiano2008CP}
Alternatively, single reference methods are sometimes used in on-the-fly nonadiabatic dynamics simulations to leverage their computational efficiency, such as those based on
configuration interaction singles (CIS), second order algebraic-diagrammatic construction [ADC(2)],
time-dependent density functional theory (TDDFT),\cite{Wang2016JPCL,Tapavicza2013PCCP} and their semiempirical analogues [such as time-dependent density functional tight binding\cite{Mitric2009JPCA}].
Spin-flip variants (spin-flip CIS and spin-flip TDDFT) have been shown to correctly describe conical intersections between the ground and excited states.\cite{Herbert2016ACR}

In this article, partly to provide a validation tool for these methods, we report a methodology for on-the-fly nonadiabatic dynamics simulations 
using newly developed analytical nuclear gradients\cite{MacLeod2015JCP,Vlaisavljevich2016JCTC} and derivative coupling\cite{Park2017JCTC}
for extended multistate multireference second-order perturbation theory (XMS-CASPT2).\cite{Finley1998CPL,Granovsky2011JCP,Shiozaki2011JCP3}
Conventionally, nonadiabatic dynamics simulations with the CASPT2 potential energy surfaces have been performed using numerical nuclear gradients
or analytical nuclear gradients for a partially contracted variant (called RS2),\cite{Celani2003JCP,Shiozaki2011JCP3}
which has limited the size of molecules to be studied owing to their computational costs. 
For instance, the fewest switches surface hopping (FSSH) simulations\cite{Tully1990JCP}
have been reported using CASPT2 numerical nuclear gradients for a reduced retinal model (ASR-M)\cite{Manathunga2016JCTC} and 2-thiouracil.\cite{Mai2016JPCL}
Other examples include the \textit{ab initio} multiple spawning simulations for ethylene and the minimal retinal model (PSB3)\cite{Levine2007ARPC,Tao2009JPCA,Virshup2009JPCB,Mori2012JPCA,Liu2016JPCB}
and the Zhu--Nakamura surface hopping simulations for thiadazole\cite{Liu2017JCP}
using RS2 analytical nuclear gradients.

The purpose of this article is to present our efficient implementation of on-the-fly XMS-CASPT2 surface hopping dynamics programs that are applicable to large molecules in condensed phases.
To realize such an implementation, 
we have substantially improved the algorithms for analytically calculating XMS-CASPT2 nuclear gradients and derivative couplings,
which are presented below.
With this improvement, one XMS-CASPT2 nuclear gradient evaluation for an anionic GFP model chromophore with CAS (4\textit{e},3\textit{o}) and cc-pVDZ
(\textit{para}-hydroxybenzilideneimidazolin-5-one or \textit{p}HBI, 21 atoms, 231 basis functions), for instance, takes
less than 2 min using 256 CPU cores.
Our program in the {\sc bagel} package\cite{bagel} is interfaced to the {\sc Newton-X}\cite{Barbatti2014WIREs} and {\sc tinker}\cite{Shi2013JCTC} packages
for nonadiabatic surface hopping dynamics and classical molecular mechanics, respectively,
to allow for simulating photodynamics in condensed phases.
As numerical examples, we present the FSSH simulations of 9H-adenine and \textit{p}HBI in aqueous solution. 

\section{Implementation}
\subsection{Improving XMS-CASPT2 nuclear gradient algorithms}

The first nuclear gradient program for fully internally contracted CASPT2 theory
has recently been reported by MacLeod and Shiozaki\cite{MacLeod2015JCP} using an automatic code generation approach,\cite{smith}
which has later been extended to multistate extensions\cite{Vlaisavljevich2016JCTC} and derivative couplings.\cite{Park2017JCTC}
These programs are implemented in the {\sc bagel} package.\cite{bagel}
In this work, we improved the algorithm substantially to make on-the-fly dynamics simulations computationally affordable, 
which is detailed in the following.

The most computationally demanding step in the XMS-CASPT2 nuclear gradient evaluation is the formation of the source terms for the so-called $Z$-vector equation.
These terms are the derivatives of the CASPT2 part of the Lagrangian $\mathcal{L}_\mathrm{PT2}$ with respect to the orbital rotation parameters $\kappa_{rs}$ and CI coefficients $c_{I}$,\cite{Celani2003JCP,MacLeod2015JCP}
\begin{align}
&Y_{rs} = \frac{\partial \mathcal{L}_\mathrm{PT2}}{\partial \kappa_{rs}}, \label{largey} \\
&y_{I} = \frac{\partial \mathcal{L}_\mathrm{PT2}}{\partial c_{I}},\label{smally}
\end{align}
where $I$ labels the Slater determinants in the active space, and $r$ and $s$ are any molecular orbitals.
The evaluation of Eq.~\eqref{smally} is particularly expensive, because the size of $y_I$ (i.e., the number of Slater determinants, or $N_\mathrm{det}$) grows factorially
with respect to the number of active orbitals.
When one evaluates Eq.~\eqref{smally},
six-index tensors calculated from the perturbation amplitudes (denoted here as $A_{ijklmn}$, $B_{ijklmn}$) and 3-particle reduced density matrix (3RDM) derivatives
or Fock-weighted 4RDM derivatives have to be contracted as
\begin{align}
y_{I} & \leftarrow
\sum_{ijklmn} A_{ijklmn} \Gamma_{ij,kl,mn}^I + \sum_{ijklmn} B_{ijklmn} \bar{\Gamma}_{ij,kl,mn}^I 
\label{yi}
\end{align}
where $i$, $j$, $k$, $l$, $m$, $n$, $o$, and $p$ label active orbitals, and $\Gamma_{ij,kl,mn}^I$ and $\bar{\Gamma}_{ij,kl,mn}^I$ are defined as
\begin{align}
\Gamma_{ij,kl,mn}^I &= \langle I | \hat{E}_{ij,kl,mn}|0\rangle,\\
\bar{\Gamma}_{ij,kl,mn}^I &= \sum_{op} \langle I | \hat{E}_{ij,kl,mn,op}|0\rangle f_{op}.
\end{align}
Here, $| 0 \rangle$ is the reference state, $f_{op}$ are Fock-matrix elements,
and $\hat{E}_{ij,kl,mn}$ and $\hat{E}_{ij,kl,mn,op}$ are the spin-free three- and four-particle excitation operators.
Previously, this step had limited the size of CASPT2 nuclear gradient calculations.

In this work, we have refactored the algorithms in order to minimize the operation counts and memory costs
for evaluating Eq.~\eqref{yi}.
In particular, storage of the 3RDM derivatives and the Fock-weighted 4RDM derivatives (whose size is $N_\mathrm{det} N_\mathrm{act}^6$) is avoided in the new algorithm, because they are
too large to be kept in (distributed) memory.
This is done by implementing a direct algorithm for evaluating Eq.~\eqref{yi}.
First, we evaluate intermediate tensors $C_{ijklmn}$ and $D_{ij}^J$,
\begin{align}
&C_{ijklmn} = A_{ijklmn} - \sum_p B_{ijklmp}f_{pn},\\
&D_{ij}^J = \sum_{kl\le mn} (2-\delta_{kl,mn})\left[ C_{ijklmn}\Gamma_{kl,mn}^J + B_{ijklmn}\bar{\Gamma}_{kl,mn}^J \right]. 
\end{align}
We then evaluate the contributions to $y_I$ in Eq.~\eqref{yi} using the following expression,
\begin{align}
y_I\leftarrow &\sum_J \sum_{ij}\langle I | \hat{E}_{ij} |J\rangle D_{ij}^J \nonumber \\
&- \sum_{ikln}\left[\Gamma^I_{il,kn}\sum_j A_{ijjlkn} + \Gamma^I_{kl,in}\sum_j A_{ijkljn}\right]
\nonumber\\
&-\sum_{ikln}\left[\bar{\Gamma}^I_{il,kn}\sum_j B_{ijjlkn} + \bar{\Gamma}^I_{kl,in}\sum_j B_{ijkljn}\right].
\end{align} 
The first term is evaluated on the fly using the algorithm used in full configuration interaction.\cite{Knowles1984CPL}
The required memory for the new algorithm is $N_\mathrm{det} N_\mathrm{act}^4$ (reduced by a factor of $N_\mathrm{act}^2$).
The program has been implemented using the automatic code generator {\sc smith3}\cite{smith}
and is efficiently parallelized using the MPI remote memory access (RMA) protocol. It is interfaced to the program package {\sc bagel}.\cite{bagel}
When the required memory is larger than the available memory, the multipassing algorithm\cite{Vlaisavljevich2016JCTC} is used in which 
10,000 elements of $y_I$ are typically computed in a single pass. 

\subsection{Interfacing QM/MM Programs}

In QM/MM simulations, systems are divided into the QM and MM regions, and the total energy $E_\mathrm{tot}$ is computed as 
\begin{align}
E_\mathrm{tot} = E_\mathrm{QM} + E_\mathrm{MM} + E_{\mathrm{QM-MM}},
\end{align}
where $E_\mathrm{QM}$ and $E_\mathrm{MM}$ are the QM and MM energies, respectively, and $E_{\mathrm{QM}-\mathrm{MM}}$ is the interaction energy between the QM and MM regions.
In this work, electrostatic embedding\cite{Aqvist1993CR} is used to evaluate $E_{\mathrm{QM}-\mathrm{MM}}$,
in which MM nuclei are represented as point charges that polarize QM wave functions (i.e., the MM part is not polarizable).
The QM--MM interaction term is an additive contribution to the one-body part of the electronic Hamiltonian as
\begin{align}
h_{rs} \leftarrow \sum_\alpha \mel** { r} { \frac{q_\alpha}{ |\mathbf{r}-\mathbf{R}_\alpha|} }{ s}.
\end{align}
where $\alpha$ labels MM nuclei, and $q_\alpha$ is the corresponding point charge located at $\mathbf{R}_\alpha$. 
The dispersive interactions between the QM and MM parts are calculated using the Lennard-Jones function.
Newton's equation of motion for the MM nuclei is integrated using the nuclear gradients of the QM/MM total energy, i.e., 
\begin{align}
- \frac{d E_\mathrm{tot}}{d \mathbf{R}_\alpha} = M_\alpha \frac {d^2 \mathbf{R}_\alpha}{d t^2},
\end{align}
where $M_\alpha$ is the mass of the MM nuclei. 
The nuclei in the QM region are propagated in a similar way. 
In the QM/MM simulations presented below, the {\sc tinker} package\cite{Shi2013JCTC} was used to generate the electrostatic charges.

\subsection{Interfacing Surface-Hopping Dynamics Programs}

In surface hopping simulations, the quantum amplitudes associated with the adiabatic electronic states are updated according to the time-dependent Schr\"odinger equation as
\begin{align}
i \frac{\partial\chi_Q}{\partial t} = \chi_Q E_Q - i \sum_P \chi_P \mathbf{v}\cdot \mathbf{d}^{QP},
\end{align}
in which $P$ and $Q$ label electronic states, $E_Q$ and $\chi_Q$ are the energy and quantum amplitude associated with electronic state $Q$, $\mathbf{v}$ is the velocity of the trajectory,
and $\mathbf{d}^{QP}$ is the derivative coupling between $Q$ and $P$.
The hop probability at time $[t, t+\Delta t]$ from $Q$ to $P$, $p_{Q\to P}$, is 
calculated according to the so-called fewest switch criteria,\cite{Tully1990JCP,HammesSchiffer1994JCP,Barbatti2011WIREs}
\begin{align}
p_{Q\to P} = \mathrm{max} \left[ \frac { 2 \mathrm{Re} \left( \chi_Q^* \chi_P \mathbf{v} \cdot \mathbf{d}^{QP} \right) \Delta t } {\chi_Q^* \chi_Q} , 0 \right].
\end{align}
When the hop probability satisfies
\begin{align}
\sum_{S=1}^{P-1} p_{Q\to S} < \zeta \le \sum_{S=1}^{P} p_{Q\to S},
\end{align}
with $\zeta$ being a random number between 0 and 1, the trajectory hops from $Q$ to $P$,
while the velocity is scaled along $\mathbf{d}^{QP}$ so that the total energy is conserved.
The hop is rejected when the energy conservation cannot be achieved by the velocity scaling.\cite{HammesSchiffer1994JCP,Jasper2002JCP,Jain2016JCTC}

The {\sc Newton-X} package\cite{Barbatti2014WIREs} is used for nonadiabatic dynamics in this work.
The communication between {\sc bagel} and {\sc Newton-X} is based on external text files.
Trajectory calculations are initiated in {\sc Newton-X}. The positions of the QM and MM nuclei are passed to {\sc bagel} in each time step.
The energies $E_{\mathrm{QM}} + E_{\mathrm{QM}-\mathrm{MM},\mathrm{ele}}$, their nuclear gradients, and derivative coupling vectors are computed by {\sc bagel},
and communicated back to {\sc Newton-X}.
The MM calculations are then performed to obtain the MM energy, dispersive interactions (i.e., $E_{\mathrm{MM}} + E_{\mathrm{QM}-\mathrm{MM},\mathrm{LJ}}$) and their nuclear gradients 
using {\sc tinker}, which are also passed to {\sc Newton-X}.
We have implemented the QM/MM calculations for the systems in which the QM and MM regions interact with each other by nonbonded interactions.
An extension of this program is planned for the systems that are connected by covalent bonds (e.g. proteins and nucleic acids) by implementing virtual link atoms.
By combining them together,  $E_{\mathrm{tot}}$ and its gradients are calculated, which are then used to integrate the equation of motion. 
This procedure is repeated till the end of the calculation.

\section{Results}

In this section, we present timing data and numerical examples to illustrate the performance of our program for on-the-fly surface hopping dynamics simulations.
All of the calculations were performed using the {\sc bagel} program package together with {\sc tinker} and {\sc Newton-X}.
The so-called ``SS-SR" contraction scheme (see Ref.~\onlinecite{Vlaisavljevich2016JCTC}) was used for internally contracted basis functions in CASPT2.
The vertical shift was set to 0.5 $E_\mathrm{h}$.
The cc-pVDZ and corresponding density-fitting basis sets were used. 

\subsection{CASPT2 nuclear gradient timings}
\begin{figure}[t]
\includegraphics[keepaspectratio,width=0.48\textwidth]{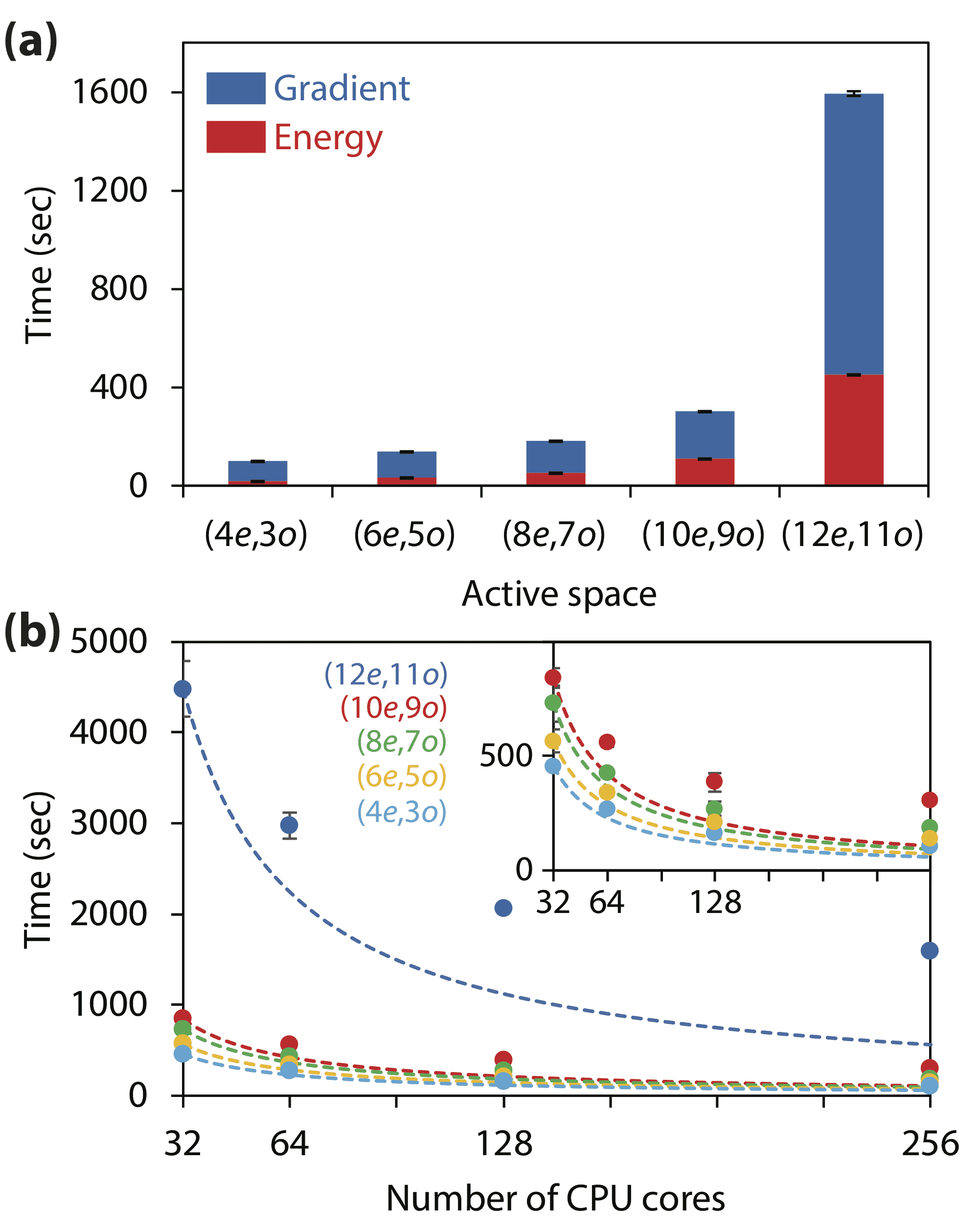}
\caption{
Computational timings for the XMS-CASPT2 nuclear gradient calculations of \textit{p}HBI using various active spaces:
(a) Wall times using sixteen nodes;
(b) their parallel scaling (strong scaling; the dashed lines are the ideal scaling).
\label{Fig_cost}}
\end{figure}

To demonstrate the performance of the new XMS-CASPT2 nuclear gradient algorithm,
the computational timings for evaluating the XMS-CASPT2 energies and nuclear gradients are shown in Fig.~\ref{Fig_cost}(a) for \textit{p}HBI using various active spaces:
CAS (4\textit{e},3\textit{o}), (6\textit{e},5\textit{o}), (8\textit{e},7\textit{o}), (10\textit{e},9\textit{o}), and (12\textit{e},11\textit{o}).
These active spaces consist of the valence $\pi$ and $\pi^*$ orbitals. The number of the atomic basis functions was 231.
Three states were averaged in the reference CASSCF calculations.
The nuclear gradient evaluation, in which the lambda and $Z$-vector equations are solved, was only twice as expensive as the energy calculation.
The timings did increase with respect to the active space size [the energy and its nuclear gradient calculations with (10\textit{e},9\textit{o})
were, for instance, about three times as expensive as those with (4\textit{e},3\textit{o})].
The new algorithm has improved the efficiency of the code, especially when large active spaces are used.
For example, the nuclear gradient calculations with (8\textit{e},7\textit{o}) and (10\textit{e},9\textit{o}) active spaces
took 261 sec and 7940 sec using the previous algorithm;
they were 131 sec and 190 sec using the current implementation.

The parallel performance is summarized in Fig.~\ref{Fig_cost}(b).
We used 2, 4, 8 and 16 nodes of our computer cluster (each node has two Intel Xeon E5-2650 2.0 GHz, total 16 CPU cores per node).
The calculations were repeated ten times, from which the averaged values were calculated.
In the (12\textit{e},11\textit{o}) case, the XMS-CASPT2 nuclear gradient evaluation with sixteen nodes (1594 sec) was 2.8 times faster than that with two nodes (4480 sec).
For small active spaces, we observed slightly better parallel performance: For instance, when the (4\textit{e},3\textit{o}) active space was used,
calculations with sixteen nodes (103 sec) were 4.4 times faster than those with two nodes (455 sec).

\subsection{Direct dynamics simulations of 9H-Adenine}
\begin{figure}[t]
\includegraphics[keepaspectratio,width=0.48\textwidth]{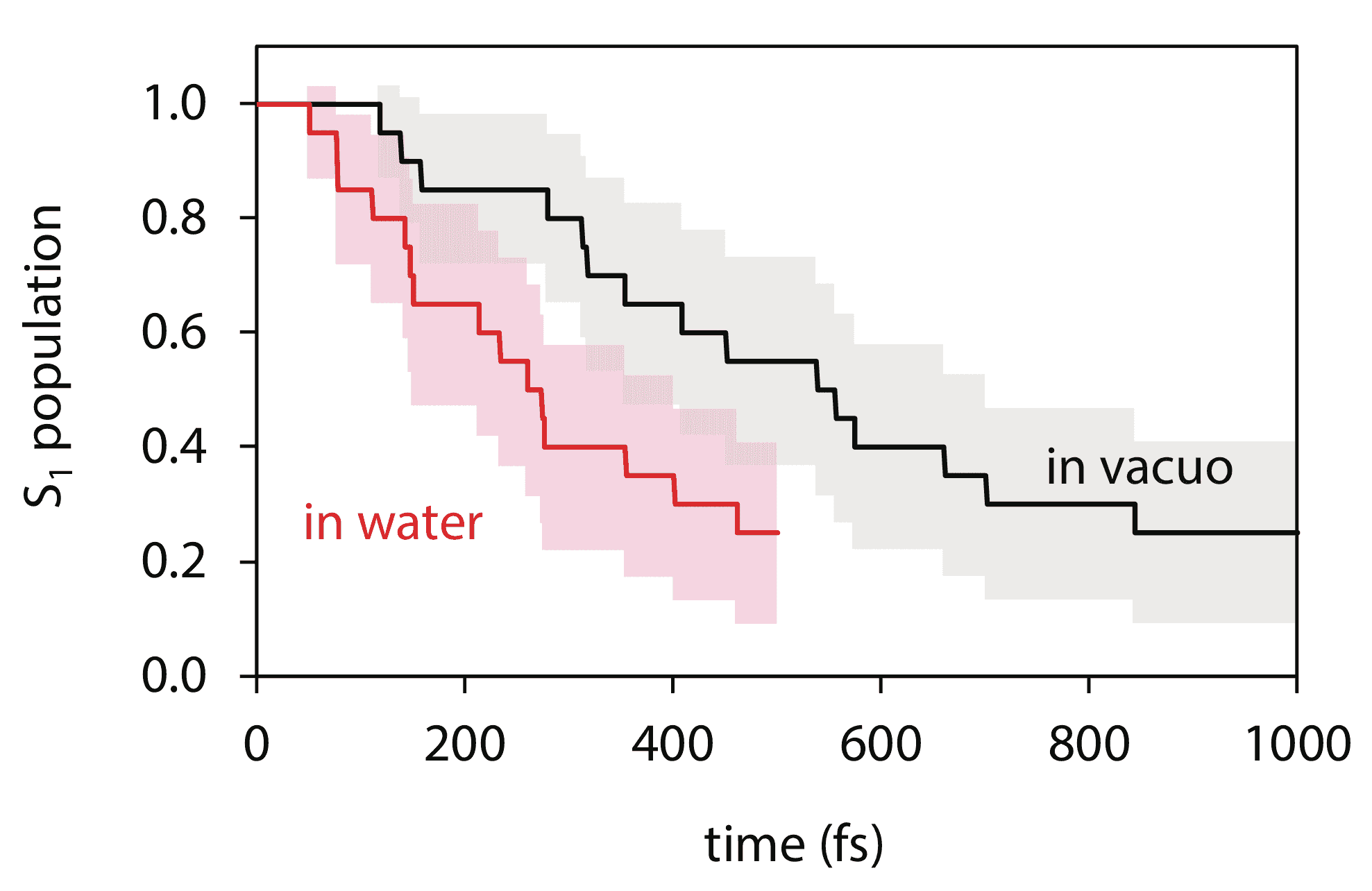}
\caption{S$_1$$\to$S$_0$ population decay dynamics of adenine in vacuo (black) and in water (red).  The bounds at confidence level 90\% are also shown.  \label{Fig_adeninepop}}
\end{figure}
\begin{figure}[t]
\includegraphics[keepaspectratio,width=0.48\textwidth]{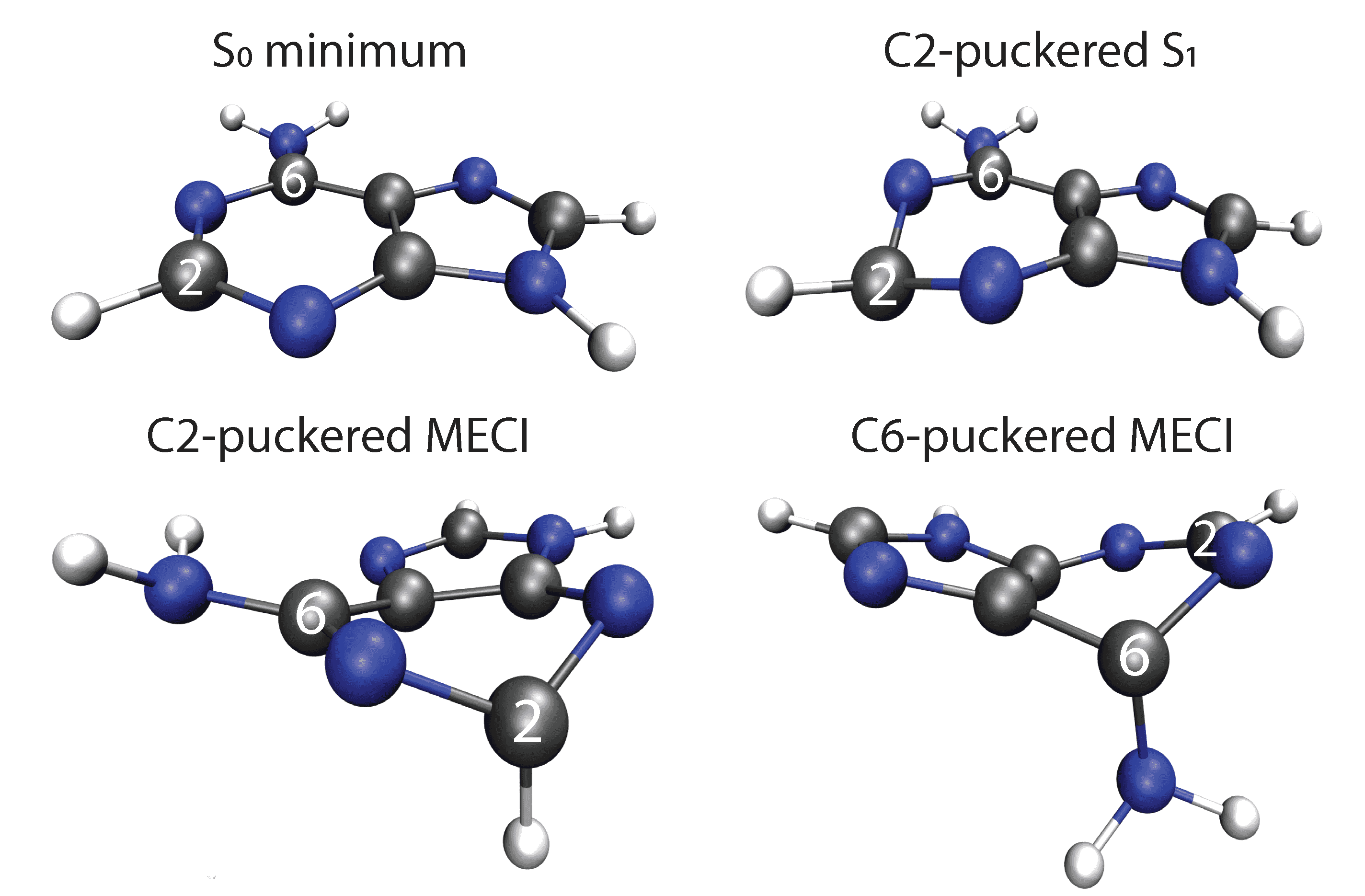}
\caption{
Molecular geometries of adenine optimized using XMS-CASPT2: Planar $\mathrm{S}_0$ minimum, C2-puckered $\mathrm{S}_1$ minimum, C2-puckered and C6-puckered MECIs.\label{Fig_adenineCI}}
\end{figure}
\begin{figure*}[t]
\includegraphics[keepaspectratio,width=0.80\textwidth]{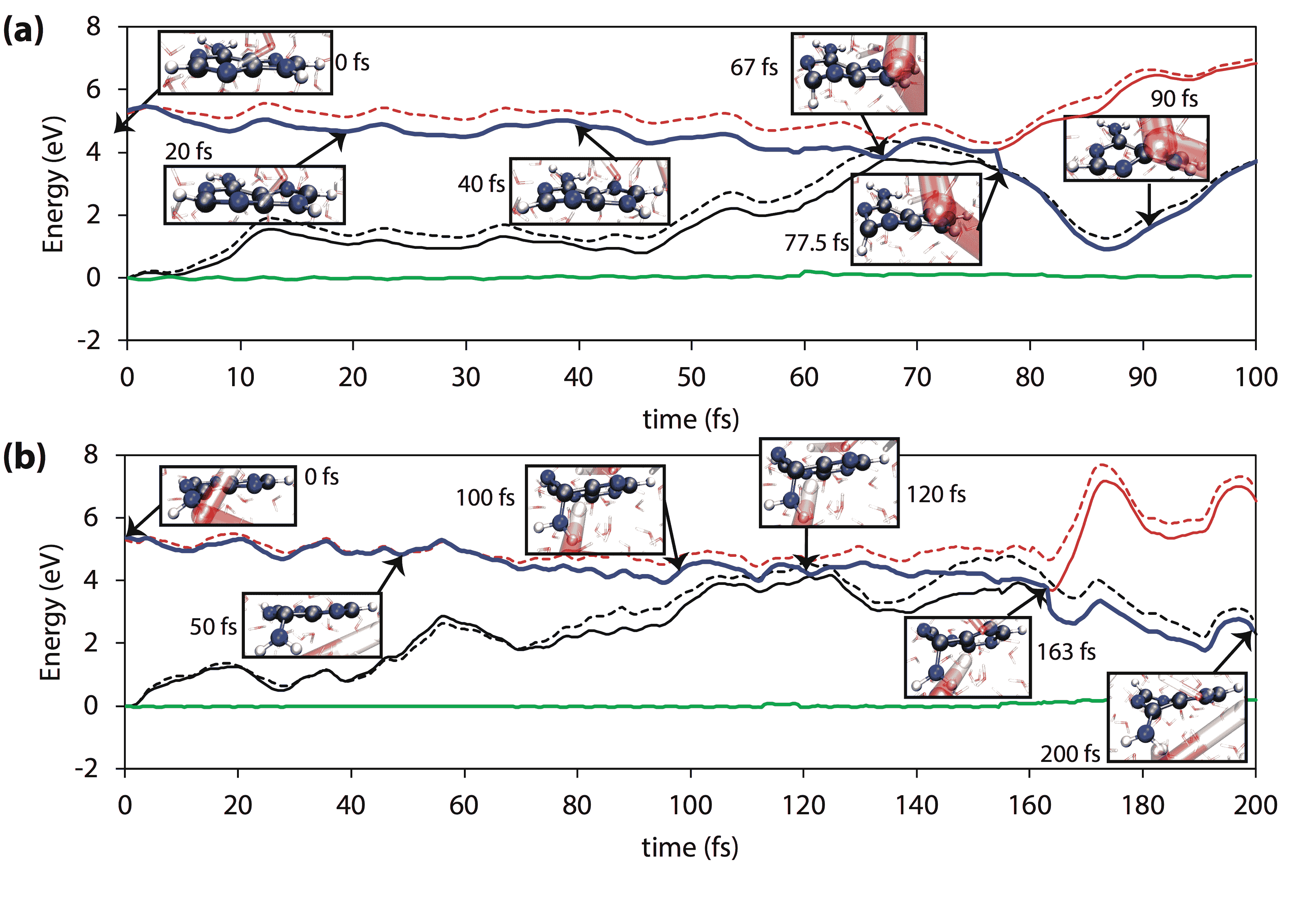}
\caption{
Selected trajectories of adenine in water, decayed through the C2-puckering (a) and C6-puckering (b) pathways.
Energies (in eV) are reported relative to the $\mathrm{S}_0$ energies at $t = 0$.
The red and black lines show the energies of the $\mathrm{S}_0$ and $\mathrm{S}_1$ states with (solid) and without (dashed) solvation contributions.
The green lines are the total energies relative to those at $t = 0$.
The blue lines represent the trajectories.
\label{Fig_adenineTraj}}
\end{figure*}

We performed the on-the-fly FSSH dynamics simulations of adenine in vacuo and water using XMS-CASPT2.
The (6\textit{e},5\textit{o}) active space was used,
which consists of two $\pi$, two $\pi^*$ orbitals and one $n$ orbital. This active space was taken from the previous MRCIS study.\cite{Barbatti2008JACS}
Five states were included in the XMS-CASPT2 calculations to account for the strong mixing among them near the conical intersections.\cite{Barbatti2012JCP}
The ground state structure of adenine was optimized using XMS-CASPT2.

The initial conditions for the gas phase trajectories were generated according to the Wigner distribution using the normal modes at the optimized geometry.
For the QM/MM trajectories, we followed the procedure in Ref.~\onlinecite{Ruckenbauer2013JPCA}:
First, 300 TIP3P waters were added to the ground state geometry using {\sc packmol}\cite{packmol} with
the spherical boundary condition [the radius was set to 13.01~{\AA} according to the water density at the standard condition (0.997 $\mathrm{g}$/$\mathrm{cm}^{3}$)];
then, the waters were equilibrated for 1 ps while the adenine molecule was forced to remain frozen.
The adenine structure was then replaced by a structure from the Wigner distribution, and waters were again equilibrated for 10 ps around the frozen adenine.
We further equilibrated the system for 200 fs using QM/MM in which MP2 is used for adenine as in Ref.~\onlinecite{Manathunga2016JCTC}.
All of the equilibration procedures were performed at $T = 300$ K.
The Lennard-Jones parameters are adopted from the OPLS all-atom force field.\cite{Jorgensen1996JACS}

Both in the gas phase and aqueous solution, we ran 20 trajectories (40 in total).
Two XMS-CASPT2 states ($\mathrm{S}_0$ and $\mathrm{S}_1$) were included in the FSSH simulations.
All of the trajectories started from the $\mathrm{S}_1$ state.
The time integration step for nuclei was set to 0.5 fs.
The quantum amplitudes were corrected for decoherence using the method by Granucci and Persico.\cite{Granucci2007JCP}
The trajectories were integrated up to 1 ps in vacuo and up to 500 fs in aqueous solution.
One compute node (16 CPU cores, Intel Xeon E5-2650 2.0 GHz) was used for XMS-CASPT2 calculations, in which
we evaluated two nuclear gradients and one derivative coupling using the analytical algorithms\cite{MacLeod2015JCP,Vlaisavljevich2016JCTC,Park2017JCTC} (about 20 minutes per time step).
The computational costs for MM gradient evaluations and trajectory integrations were negligible compared to those for XMS-CASPT2. 
Trajectories were discarded when the total energy changed abruptly more than 1.0 eV due to unstable CASSCF as is commonly done.\cite{Fingerhut2013JPCL,Mai2016JPCL}

The simulated population decay patterns are shown in Fig.~\ref{Fig_adeninepop}. When fitted to a single-exponential function,
the time constants are 760 fs and 386 fs in the gas phase and in aqueous solution, respectively.
At confidence interval 90\%, their ranges are 478--1250~fs (in vacuo) and 237--653~fs (in water)
[these time constants are obtained from the upper and lower bounds of the population decay patterns that are calculated using Eq.~(6) of Ref.~\onlinecite{Plasser2014JCTC}].
The experimental results are about 1.2 ps\cite{Satzger2006PNAS} and 180 fs,\cite{Cohen2003JACS}
respectively.
The estimated time constant in the gas phase is slightly shorter than the experimental value, partly because the excitation to $\mathrm{S}_2$ is not included in our simulations
(which contributes about 30\% of all the excitations), which additionally requires about 40 fs to decay into $\mathrm{S}_1$.\cite{Barbatti2008JACS}
The rapid decay to the ground state in the presence of the solvent molecules is qualitatively reproduced by the QM/MM simulations.

The two major conical intersections, called C2-puckered and C6-puckered, are responsible for the nonadiabatic decays.\cite{Barbatti2008JACS,Barbatti2012JCP}
The minimum energy conical intersection (MECI) geometries are shown in Fig.~\ref{Fig_adenineCI}.
The energies at the MECIs are 4.00 eV (C2-puckered) and 4.42 eV (C6-puckered) above the ground state energy at the optimized geometry,
while the energy of the $\mathrm{S}_1$ state at the Franck--Condon point is 5.52 eV.
The energy of the $\mathrm{S}_1$ state at the C2-puckered $\mathrm{S}_1$ minimum was 4.64 eV.
This means that both MECIs are thermally accessible on the XMS-CASPT2 surface.

The geometries at which hops occur can be likewise classified into the C2-puckered and C6-puckered pathways.
In the previous works, the nonadiabatic decays in vacuo are reported to be
dominated by C2-puckering (79\% in the gas phase) with MRCIS\cite{Barbatti2012JCP} and by C6-puckering (95\% in the gas phase\cite{Fabiano2008JPCA} and 90\% in aqueous solution\cite{Lan2011CPC}) with OM2/MRCI,
which disagree with one another.
In our simulations, 64\% and 80\% of the trajectories that experienced a hop decayed through the C2-puckering pathway in the gas phase and in aqueous solution, respectively.
The ADC(2) results reported in Refs.~\onlinecite{Plasser2014JCTC} and \onlinecite{Barbatti2014JACS}
[52\%--54\% (gas phase) and 90\% (adenine--water cluster) reached near the C2-puckering conical intersection] are also consistent with our results.
The discrepancy of about 10\% can be in part explained by the small ensemble size in our simulation.

To understand the mechanism of faster nonadiabatic decays in water, the selected trajectories are shown in Fig.~\ref{Fig_adenineTraj}.  
Our trajectories show that the stabilization of the potential energy surfaces due to solvation make the conical intersections more accessible, leading to faster decay.
To see the changes of the energies due to solvation, we re-calculated the XMS-CASPT2 energies at each snapshot without solvation contributions,
which are also shown in Fig.~\ref{Fig_adenineTraj}.
In particular, the out-of-plane displacements of the hydrogen atom (C2-puckering) or the amino group (C6-puckering) that lead to the nonadiabatic decays are stabilized by the interaction between adenine and water.
The energies of the MECIs in water are 1.63 eV (C2-puckering) and 0.80 eV (C6-puckering) below the energy of the $\mathrm{S}_1$ state at the Franck--Condon point, respectively,
which also implies that both conical intersections are accessible in water.

\subsection{Direct dynamics simulations of \textit{p}HBI}
\begin{figure}[t]
\includegraphics[keepaspectratio,width=0.48\textwidth]{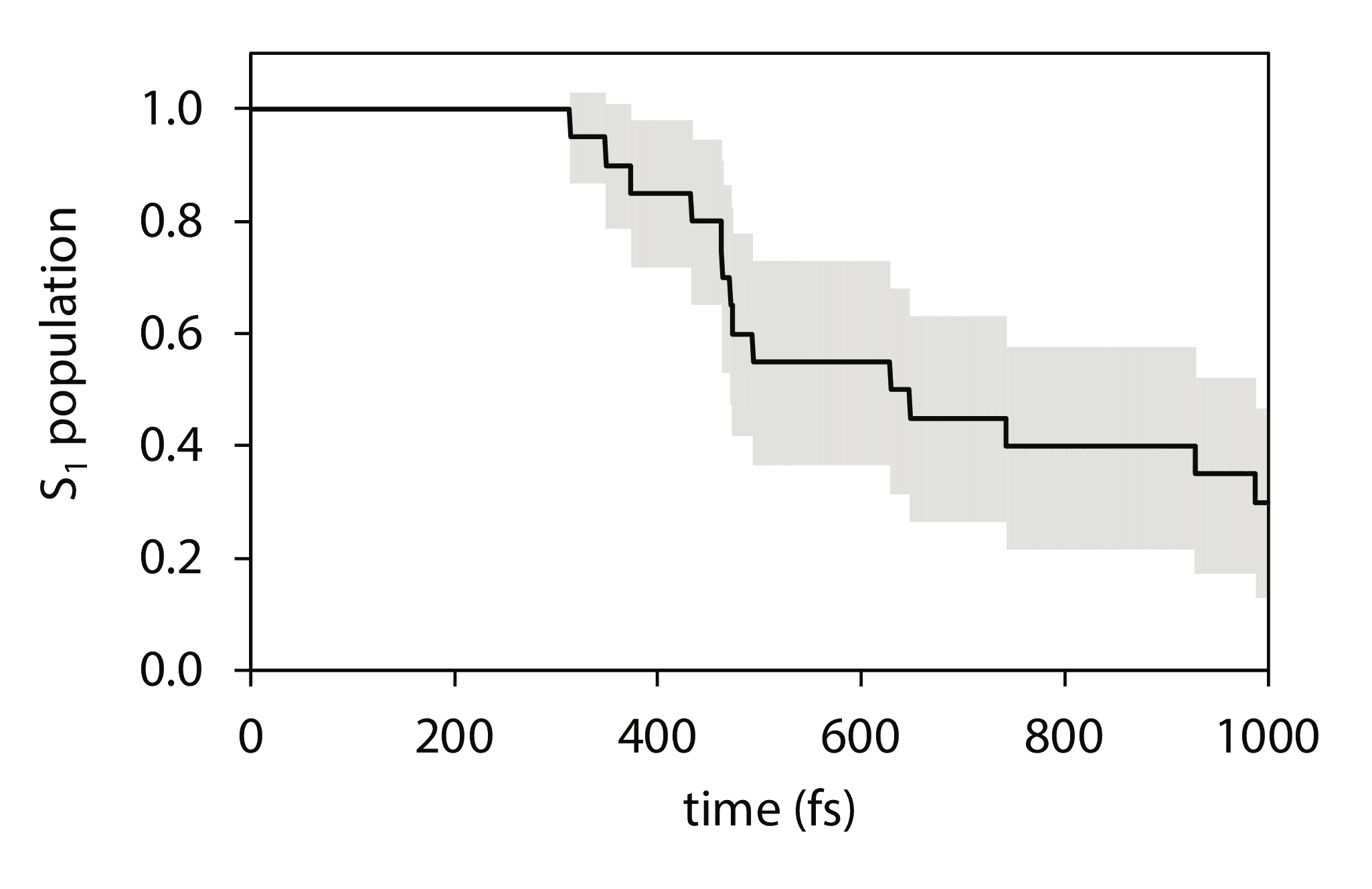}
\caption{S$_1$$\to$S$_0$ population decay dynamics of \textit{p}HBI in water.  The bounds at confidence level 90\% are also shown.  \label{Fig_phbipop}}
\end{figure}
\begin{figure}[t]
\includegraphics[keepaspectratio,width=0.48\textwidth]{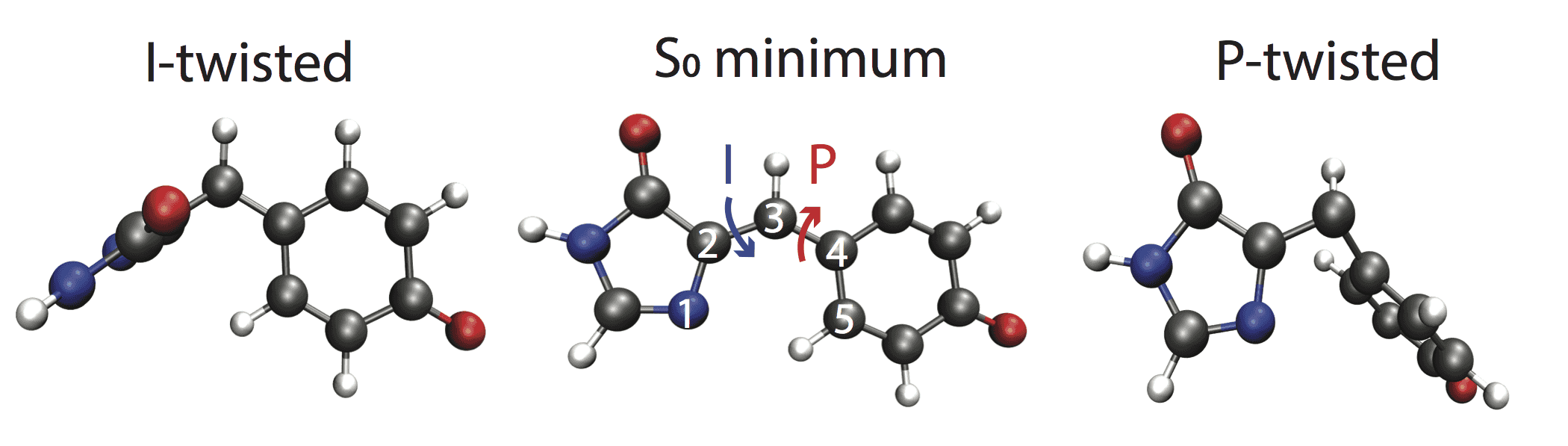}
\caption{
Molecular geometries of \textit{p}HBI optimized using XMS-CASPT2: Planar $\mathrm{S}_0$ minimum, I- and P- twisted MECIs.
\label{Fig_phbiCI}}
\end{figure}
\begin{figure}[t]
\includegraphics[keepaspectratio,width=0.48\textwidth]{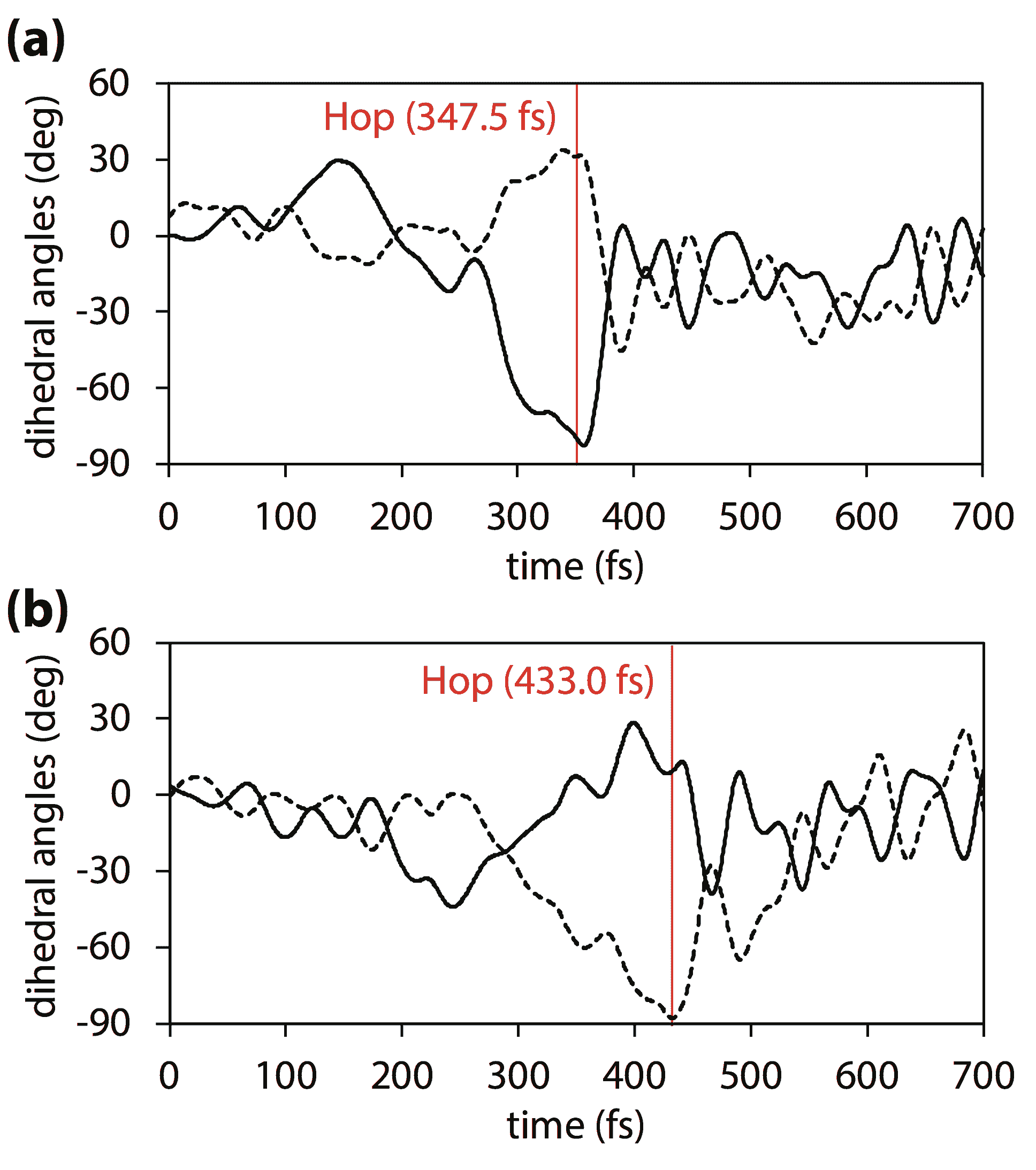}
\caption{
P-(dashed) and I-(solid) dihedral angles of \text{p}HBI in selected trajectories, decayed through the I-channel (a) and the P-channel (b).
\label{Fig_phbi_dihedral}}
\end{figure}
\begin{figure*}[t]
\includegraphics[keepaspectratio,width=0.80\textwidth]{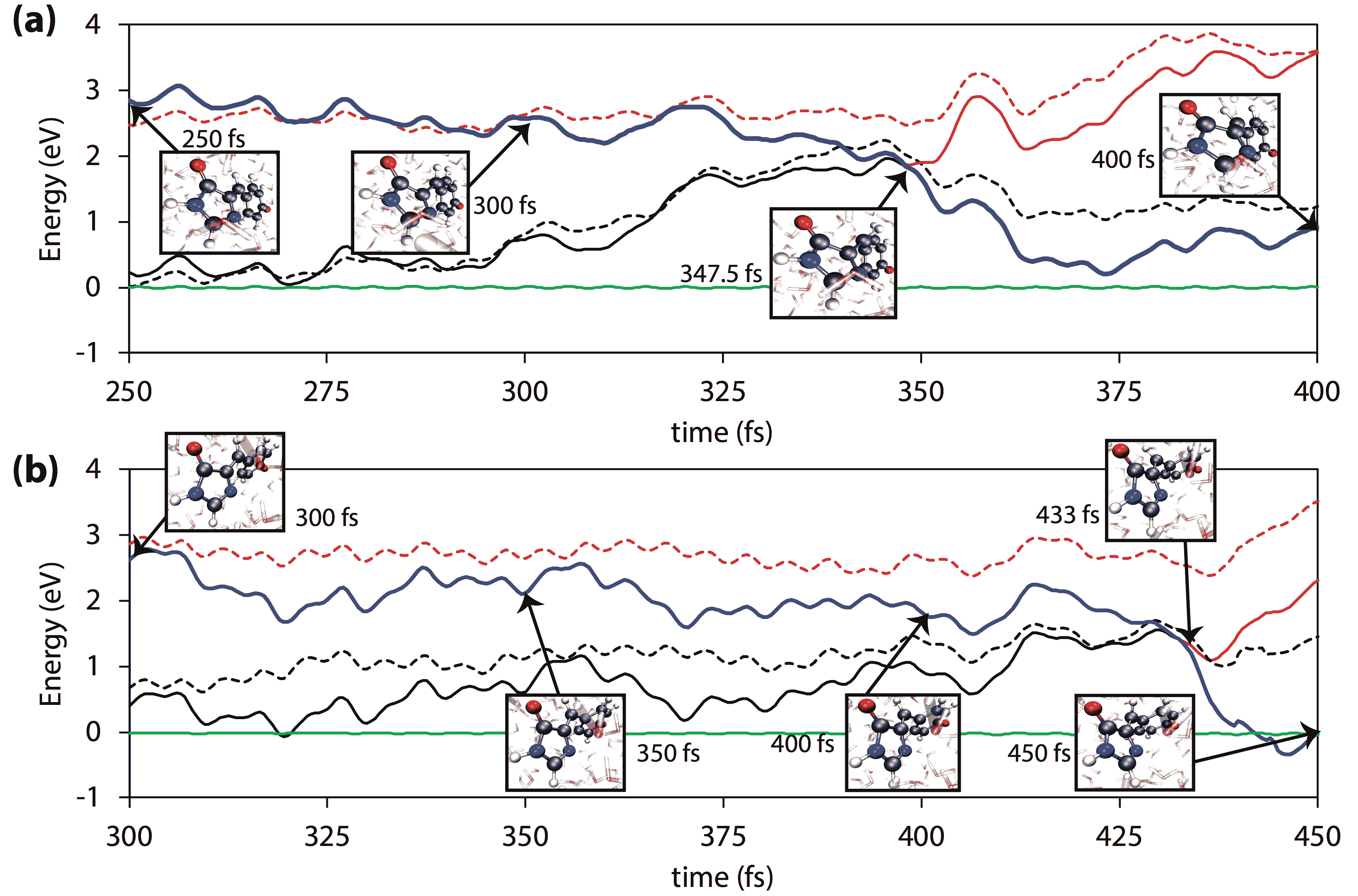}
\caption{
Selected trajectories of \textit{p}HBI in water, decayed through the I-channel (a) and the P-channel (b).
Energies (in eV) are reported relative to the $\mathrm{S}_0$ energies at $t = 0$.
The red and black lines show the energies of the $\mathrm{S}_0$ and $\mathrm{S}_1$ states with (solid) and without (dashed) solvation contributions. 
The green lines are the total energies relative to those at $t = 0$.
The blue lines represent the trajectories.
\label{Fig_pHBITraj}}
\end{figure*}

We performed the on-the-fly FSSH dynamics simulations of an anionic GFP model chromophore, \textit{p}HBI, in water using XMS-CASPT2.
The (4\textit{e},3\textit{o}) active space with three-state averaging was used in the XMS-CASPT2 calculations.
The previous single-point CASPT2 studies\cite{Olsen2008JACS,Olsen2009JCP} have shown that this (4\textit{e},3\textit{o}) active space 
describes in a balanced manner the neutral and charge transfer states associated with twisting of the chromophore.
The same protocol to generate the initial conditions described in the previous section was used.
The Lennard-Jones parameters were taken from Ref.~\onlinecite{Reuter2002JPCB}, which are defined in the CHARMM22 force field.\cite{MacKerell1998JPCB}
We included the two lowest singlet electronic states in the FSSH simulations.
All of the trajectories were initiated from the first excited state. We ran 20 trajectories in total.
The time integration step for nuclei was set to 0.5 fs. All of the trajectories were integrated up to 1 ps.
Eight compute nodes on Cray XE6, each of which has two AMD Interlagos Opteron 6281 2.5 GHz CPUs (16 cores each, total of 256 CPU cores),
were used for XMS-CASPT2 calculations.
Two nuclear gradients and one derivative coupling vector were evaluated at each time step, which took about 10 min.
Similarly to the case of adenine, the computational costs for the trajectory propagation were negligible compared to those for XMS-CASPT2.

The simulated population decay pattern is shown in Fig.~\ref{Fig_phbipop}. Within 1 ps, 70\% of the trajectories decayed down to the ground state.
The experimental fluorescence decay constants have been reported to be 0.21 ps (51\%) and 1.1 ps (49\%).\cite{Mandal2004JPCB}
The nonadiabatic decays from $\mathrm{S}_1$ to $\mathrm{S}_0$ occurred after the twisting of the bridge bonds;
two bridge bonds that are connected to the imidazolinone (I-channel) and phenoxy (P-channel) rings can be twisted.
The molecular structures of the conical intersections that correspond to these pathways are shown in Fig.~\ref{Fig_phbiCI}.
The previous CASSCF studies have suggested decays through the P-channel.\cite{Olsen2010JACS,Zhao2014JCP}
In our XMS-CASPT2 trajectory simulations, the major pathway was found to be the I-channel (71\% of the trajectories that experienced a hop decayed through the I-channel).
The energies at the I-twisted and P-twisted MECIs are 0.30 eV lower and 0.22 eV higher than the $\mathrm{S}_1$ energy at the Franck--Condon point, respectively,
making the P-twisted conical intersections thermally less reachable than the I-twisted conical intersections.

The bridge dihedral angles in selected trajectories are shown in Fig.~\ref{Fig_phbi_dihedral}.
The P and I dihedral angles are defined by the atoms C2--C3--C4--C5 and N1--C2--C3--C4, respectively, as shown in Fig.~\ref{Fig_phbiCI}.
The chromophore retains its planar geometry for several hundreds of femtoseconds after photoexcitation;
next, one of the bridge bonds suddenly twists, and the chromophore reaches the conical intersection close to the twisted geometry;
subsequently, shortly after the nonadiabatic decay to the ground state, the chromophore restores its planar geometry.
Although the decay pathway is not the same as those in the previous studies, the sudden twists are observed in both pathways,
which is consistent with the insights obtained by the previous studies using CASSCF\cite{Olsen2010JACS,Zhao2014JCP}
or fitted potential energy surfaces.\cite{Jonasson2011JCTC,Park2016JACS}

Finally, to investigate how solvation affects the nonadiabatic decays in water, we re-calculated XMS-CASPT2 energy at each snapshot without solvation contributions
in the same trajectories (see Fig.~\ref{Fig_pHBITraj}).
As in adenine, the conical intersections are stabilized by solvation, and both conical intersections become thermally reachable.
This is confirmed by the energies of the MECIs in water that correspond to the I and P channels
(0.46 eV and 0.72 eV lower than the energy of the $\mathrm{S}_1$ state at the Franck--Condon point).
The $\mathrm{S}_1$--$\mathrm{S}_0$ energy gaps become much smaller at the twisted geometries in water (by about 1.0~eV) than in the gas phase.
This is due to the significant charge transfers due to excitation as also discussed in Refs. \onlinecite{Olsen2009JCP} and \onlinecite{Martin2004JACS}.
The changes of the Mulliken populations at the MECIs on the phenoxy ring (i.e. $q_{\mathrm{S}_1} - q_{\mathrm{S}_0}$), computed using the XMS-CASPT2 relaxed density matrices,
are $-0.46$ at the I-twisted MECI and $0.66$ at the P-twisted MECI.
In addition, the potential energy surfaces near the MECIs are found to be less sloped in water, to which the rapid nonadiabatic decays can be ascribed.
The energies, Mulliken populations, and potential energy surfaces near the MECIs are compiled in the Supporting Information.

\section{Conclusions}

In summary, we have developed a computational tool for on-the-fly nonadiabatic dynamics simulations in the gas and condensed phases using analytical nuclear gradients and derivative couplings
for XMS-CASPT2 that account for both static and dynamical electron correlation.
We presented the simulations for the photodynamics of adenine and GFP model chromophore,
which had not been tractable prior to our development.
The {\sc bagel} program package, including the interface to the nonadiabatic dynamics program {\sc Newton-X}, is distributed under the GNU General Public License,
which allows users to freely download, modify, and redistribute it under the same terms.

\section{Acknowledgments}
The authors thank Ryan Reynolds for executing some of the trajectory calculations.
This work has been supported by the Air Force Office of Scientific Research Young Investigator Program (Grant No.~FA9550-15-1-0031)
and the ERDC DSRC high-performance computing resources (AFOSR40403702).
The development of the program infrastructure has been in part supported by National Science Foundation [ACI-1550481 (JWP) and CHE-1351598 (TS)].
T.S. is an Alfred P. Sloan Research Fellow.

\end{document}